\documentstyle[12pt]{article}
\renewcommand{\baselinestretch}{1.7}

\def\lsim{\mathrel{\rlap{\raise 2.5pt \hbox{$<$}}\lower 2.5pt}}
\def\gsim{\mathrel{\rlap{\raise 2.5pt \hbox{$>$}}\lower 2.5pt}}
\begin{document}
\thispagestyle{empty}
\vspace{-3mm}
\begin{small}
\begin{flushright}
BUTP-96/6, KL-TH~96/4, hep-ph/9601372 \\
\end{flushright}
\end{small}
\begin{center}
{\bf{\Large The Non-Minimal Supersymmetric \\
Standard Model at Large $\tan\beta$}}
\vskip 0.5cm
B. Ananthanarayan\\[-2mm]
\begin{small}
Institut f\"ur Theoretische Physik, Universit\"at Bern,\\[-2mm]
CH 3012, Bern, Switzerland\\[-2mm]
\vskip 0.2cm
\end{small}
and
\vskip 0.2cm
P. N. Pandita\\[-2mm]
\begin{small}
Universit\"at Kaiserslautern, Fachbereich Physik,\\[-2mm]
Erwin--Schr\"odinger--Strasse, D-67663 Kaiserslautern, Germany\\[-2mm]
and\\[-2mm]
Department of Physics, North Eastern Hill University,\\[-2mm]
Shillong 793 003, India$^{*}$\\[-2mm]
\end{small}
\end{center}
\begin{abstract}
We present a comprehensive analysis  of the 
Non-Minimal Supersymmetric
Standard Model (NMSSM) 
for large values of
$\tan\beta$, the ratio of the vacuum expectation values of the 
two Higgs doublets,
which arise when we  impose the constraint of the unification
of Yukawa couplings in the model.
In this limit we show that the vacuum 
expectation value of the singlet is forced to be large, 
of the order of $10$ TeV.
The singlet decouples from the lightest
CP-even neutral Higgs boson and
the neutralinos.
We compare our results with the corresponding particle spectrum
of the Minimal Supersymmetric Standard Model in the same limit.
With the exception of the lightest Higgs boson, the
particle spectrum in the model turns out to be heavy.  The 
Higgs boson mass, after the inclusion of radiative corrections, 
is found to be in the neighbourhood of $\sim 130$ GeV.
\end{abstract}
\noindent{\underline{\hspace{11.6cm}}}\\[-2mm]
* Permanent address 


\newpage
\setcounter{equation}{0}
\section{Introduction.}

\bigskip

The recent determination of the coupling constants of the
$SU(3)_C\times SU(2)_L\times U(1)_Y$ model of strong and
electroweak interactions is compatible with the supersymmetric
unification~\cite{Gauge} of these couplings with a supersymmetry~\cite{HPN}
breaking scale of the order of $1$ TeV.  In
supersymmetric (SUSY) models, at least two higgs doublets
($H_1$ and $H_2$) with opposite hypercharge are required to give
masses to the up- and down- type quarks (and charged leptons),
and to cancel gauge anomalies.  The ratio of the vacuum expection values
of the neutral components ($<H_1^0>=v_1$ and $<H_2^0>=v_2$)
of these two Higgs doublets, $\tan\beta\equiv {v_2\over v_1}$,
is a crucial parameter for the predictions from supersymmetric grand unified
theories.  
One particularly predictive framework
is based on the assumption that the heaviest generation
fermions lie in a unique {\bf 16}-dimensional representation
of the unifying gauge group $SO(10)$,  with the Higgs doublets
in a {\bf 10}-dimensional representation of the group[3]. 
Furthermore, if one makes the additional assumption that the
fermion masses are generated by a single complex {\bf 10}-plet
of $SO(10)$ through a $h. {\bf 16}.{\bf 16}.{\bf 10}$ term in
the superpotential at the unification scale $M_X$,  determined
from gauge coupling unification, one would have the prediction~\cite{ALS1}:
\begin{equation}\label{yukawaunification}
h_t=
h_b=h_\tau=h
\end{equation}
where $h_{t,b,\tau}$ are the Yukawa couplings of $t,\ b$ and $\tau$.
The coupled system of gauge and Yukawa couplings are then evolved from 
$M_X$ down to the weak scale, and $\tan\beta$ is determined from the
accurately measured value of $m_\tau=1.78\ $ GeV.  When $h$ is chosen in such 
a manner as to yield a value for $m_b(m_b)$ in its observed range of
$4.25\pm 0.10$ GeV~\cite{GL}, a rather good prediction for the top-quark
mass is obtained, which, with the present central
value of $\alpha_S(M_Z)=0.12$, lies in the range favored by the
experimental data~\cite{CDF}.  In such a situation, $\tan\beta$
is found to saturate what is considered to be a theoretical upper bound
on its values of $m_t/m_b$, and the Yukawa coupling $h$ is found
to come out to be rather large $O(1-3)$, with a certain insensitivity
to the exact value, since it is near a fixed point of its evolution.

At the weak scale, the minimal particle content of such a grand unified theory
is that of the minimal supersymmetric standard model~\cite{HPN}, with each
bosonic (fermionic) degree of freedom of the standard model complemented
by fermionic (bosonic) one, apart from having two Higgs doublets
as mentioned earlier.  However, it does not involve in any great detail
the remaining aspects of the embedding of the standard model into a 
supersymmetric grand unified framework.  With the particle content of
the MSSM and an additional discrete symmetry, called R-parity~\cite{HPN}, 
which forbids
couplings that can lead to rapid nucleon decay, it is possible to construct
a self-consistent and successful framework.  
In particular, the Higgs sector of the MSSM contains five physical
degrees of freedom: two CP even ($h^0$ and $H^0$) and
one CP-odd ($A^0$) neutral and complex charged Higgs bosons ($H^\pm$).
The spectrum of the Higgs bosons is strongly affected by the 
grand unified theory assumptions such as the ones described above.
For instance, the mass of the lightest higgs boson $m_{h^0}$, which
is related through the D-term in the potential to the mass of the Z-boson 
($m_Z$), after inclusion of radiative corrections is found to be 
$\stackrel{<}{\sim}
140$ GeV~\cite{ALS2}.  Furthermore, supersymmetry breaking is understood
to arise from embedding the MSSM into a supergravity framework and
writing down all the possible soft supersymmetry breaking terms
consistent with gauge and discrete symmetries that define the model.
It is ususally assumed that many of the  parameters describing these
terms are in fact equal at the unification scale in order to have a
predictive framework, which is  motivated by such arguments as the weak
principle of equivalence when applied to coupling of the SUSY  breaking
hidden sector to the known sector via gravitation.  Such universal
boundary conditions have had to be modified somewhat in order to
avoid fine tuning and to minimize the effects of possible radiative
corrections to the b-quark mass induced through the supersymmetry
breaking sector of the theory at the one-loop 
level~\cite{ALS2,OP,Anderson,Matalliotakis}.

In $SU(5)$ type unification where $\tan\beta$ is free, the region
$\tan\beta\simeq 1$ is also a region which is favoured for the unification
of  the b-quark and $\tau$-lepton Yukawa couplings from the observed
data~\cite{Arason}.
One crucial difference between the two extremes is that for large values
of $\tan\beta$, the Yukawa couplings of the b-quark (and
that of the $\tau$-lepton)
always remain comparable to that of the top-quark, with the observed
hierachy in their masses arising from the large value of $\tan\beta$;
when $\tan\beta\simeq 1$, the Yukawa couplings of the b-quark and
that of the $\tau$-lepton are much smaller than that of the top-quark.

Despite its many successes, it may be premature to confine our attention
only to the MSSM, especially because of the presence of dimensionful
Higgs bilinear parameter $\mu$ in the superpotential.
An alternative to the MSSM that has been widely considered is the
Non-Minimal Supersymmetric Standard Model (NMSSM) , where the
particle content of the Minimal Supersymmetric Standard Model (MSSM)~
is extended~ \cite{Fayet} by the addition of a gauge singlet chiral superfield $S$, and
one in which dimensionful couplings are eliminated through the introduction
of a discrete $Z_3$ symmetry.
While the explicit coupling $\mu H_1 H_2$ is forbidden, an effective
$\mu$ term is generated by the vacuum expectation value $<S>
(\equiv s)$ of the singlet Higgs field.

The  NMSSM is characterised by the
superpotential~ \cite{Ellis}
\begin{equation}\label{superpotential}
W=h_t Q\cdot H_2 t^c_R+h_b Q\cdot H_1b^c_R+h_\tau L\cdot H_1\tau^c_R+
\lambda S H_1.H_2+\frac{1}{3}kS^3,
\end{equation}
where we have written the interactions of only the heaviest
generation and the Higgs sector of the theory.  In addition, one
must add to the potential obtained from
eq.(\ref{superpotential}), the most general
soft supersymmetry breaking terms.  These are:
\begin{eqnarray}
& \displaystyle
(h_tA_t \tilde{Q}\cdot H_2 \tilde{t^c_R}+
h_b \tilde{Q}. H_1 \tilde{b^c_R} +
h_\tau A_\tau \tilde{L} \cdot H_1\tilde{\tau^c_R} +
\lambda A_\lambda H_1\cdot H_2 S +
{{1}\over{3}}k A_k S^3)+ {\rm h. c.} & \nonumber \\
& \displaystyle +m_{H_1}^2|H_1|^2+m_{H_2}^2|H_2|^2+m_S^2|S|^2
+m_{\tilde{Q}}^2|{\tilde{Q}}|^2+
m_{\tilde{t}}^2|{\tilde{t}}^c_R|^2
+m_{\tilde{b}}^2|{\tilde{b}}^c_R|^2
+m_{\tilde{\tau}}^2|{\tilde{\tau}}^c_R|^2 & \nonumber \\
& \label{softterms}
\end{eqnarray}
in the standard notation. In addition, there are soft SUSY breaking mass
terms for $U(1)_Y$, $SU(2)_L$ and $SU(3)_C$ gauginos 
($\lambda^{\prime};~ 
\lambda^a,~ a = 1,~ 2,~ 3;~ \lambda^i; ~i = 1, ....,8$), which we shall
denote by $M_1$, $M_2$ and $M_3$, respectively. 
We note that if $k=0$, the Lagrangian 
obtained from (1.2) 
has a global $U(1)$ symmetry
corresponding to $N\rightarrow N e^{i\theta}$, $H_1 H_2 \rightarrow
H_1 H_2 e^{-2i\theta}$, which is broken by the vacuum expectation
values (VEVs) of the Higgs fields.  In order to avoid axion associated
with this symmetry we require $k\neq 0$.

In this paper we present a study of the particle spectrum of 
the NMSSM, characterized
by eq.(\ref{superpotential}) and eq.(\ref{softterms}), for large values of 
$\tan\beta$~\cite{AP}
in detail, and compare it with the corresponding
spectrum in the MSSM.  We carry out a renormalization group
analysis of this model with universal boundary conditions and analyze
the RG improved tree-level potential at the scale $Q_0$.  The cut-off
scale for the renormalization group evolution is chosen to be the
geometric mean of the scalar top quark masses which is roughly equal
to the corresponding mean of the scalar b-quark masses as well, since
the evolution of the Yukawa couplings of the t- and b-quarks are equal
upto their hypercharges and the relatively minor contribution of the
$\tau$-lepton Yukawa coupling [we note here that this feature is
crucial in justifying our use of the tree-level potential since
with large $\tan\beta$, the Yukawa coupling of the b-quark is large,
influencing
the contributions of the large logarithms in the one-loop potential].
Whereas in the MSSM the parameters $\mu$ and $B$ (the soft SUSY breaking
parameter corresponding to the bilinear term in the superpotential
of the MSSM) do not enter into the evolution of the other parameters
of the model at the one-loop level, the situation encountered here is 
drastically different with a systematic search in the parameter space
having to be performed with all parameters coupled from the outset.
Our analysis of the minimization conditions that ensure a vacuum
give rise to severe fine tuning problems that are worse in the NMSSM
in comparison to those in the MSSM.  The problem is further compounded
by our having to satisy three minimization conditions in contrast to the 
two that occur in the MSSM.  In studies of the model where $\tan\beta$
is  a free and adjustable parameter, the tuning of parameters is possible in
order to meet all the requisite criteria, viz., minimization conditions,
requirement that the vacuum preserve electric charge and color, etc.
However in the present case where $\tan\beta$ is fixed and large, what
we find is a highly correlated system.

The plan of the paper is as follows:  In Section 2, we review the
basic framework of the MSSM at large $\tan\beta$ so as to set up
the stage for our analysis of the NMSSM in the same limit, and
thereby  enable us to compare and contrast the two models.  In Section 3
a detailed analysis of the NMSSM in this limit is presented.
An important conclusion of our analysis is that the vacuum expectation
value of the singlet is forced to be large in this limit.  In Section 4
we present a numerical study of the model and the conclusions.

\setcounter{equation}{0}
\section{MSSM at Large $\tan\beta$}

As mentioned in the introduction, in the minimal supersymmetric
extension of the standard model, the ratio of the vacuum expectation
values of the two Higgs doublets, $\tan\beta$, is an important parameter
that determines the spectrum in an essential way.  The unification
of the third family Yukawa couplings, eq.(\ref{yukawaunification}), 
leads to $\tan\beta$
being determined in the MSSM, thus, reducing the  parameter space  
significantly.  The analysis of the MSSM starts with the scalar potential,
where the parameters are evolved according to the one-loop renormalization
group equations from $M_X$ to the low energy scale $Q_0$~\cite{ALS2}.
At tree-level it is possible to analytically find the minimum of the 
potential, and we simply quote the results here~\cite{HK}. We start with the 
definitions:
\begin{eqnarray*}
\mu_1^2=m_{H_1}^2+\mu^2, & \mu_2^2=m_{H_2}^2+\mu^2, & \mu_3^2=\mu B
\end{eqnarray*}
where $m_{H_i}^2, \ i=1,2$ are the soft-supersymmetry breaking mass
parameters for the Higgs fields,  and $B$ is the soft-supersymmetry breaking
bilinear parameter corresponding to  the $\mu H_1 H_2$ term in the
superpotential of MSSM.  In order that the minimum
of the potential break $SU(2)\times U(1)$ gauge symmetry to $U(1)_{em}$, 
we must have $\mu_1^2 \mu_2^2 < \mu_3^4$, whereas to prevent the potential from
being unbounded from below we require $\mu_1^2+\mu_2^2 \geq 2 |\mu_3|^2$
(note that the left-hand side of this is nothing but the square
of the mass of the CP-odd Higgs boson, $m_A^2$).  
These requirements tend to favor a
situation when only one of $\mu_1^2$ or $\mu_2^2$ is negative, or where both
are positive and one is somewhat smaller than the other.
From here we see that radiative electroweak symmetry breaking
via the effects of $h_t$ may be implemented  in the MSSM in general~\cite{IR},
as it drives $\mu_2^2$ to smaller (possibly negative) values in
relation to $\mu_1^2$,  while $\mu_3^2$ can be easily set negative.

In the attractive scenario of radiative electroweak symmetry breaking,
the mass parameters $\mu_1^2$ and $\mu_2^2$ start out at the GUT
scale with a universal positive value
\begin{eqnarray*}
\mu_1^2=\mu_2^2=m_0^2+\mu^2,
\end{eqnarray*}
where $m_0$ is the universal scalar mass.  Thus the symmetry is
not broken at this scale.  However, in the RG evolution to the
electroweak scale, the large Yukawa coupling of the top quark to
$H_2$, which gives the top its mass, also drives the mass-squared
parameter $\mu_2^2$ of $H_2$ negative (competing against the
QCD coupling), while the absence of a large Yukawa in the down
sector keeps the mass squared of $H_1$ positive.  The above conditions
are easily satisfied for a large range of initial conditions
if $h \sim O(1)$, resulting in a very natural picture of
electroweak symmetry breaking.  This picture is essentially
lost in the large $\tan\beta$ scenario, for the reasons discussed
below.

Firstly, since all the Yukawa couplings are comparable, the two higgs
doublets tend to run in the same way, so  either both $\mu_1^2$ and
$\mu^2_2$ stay positive at the electroweak scale and the symmetry
does not break, or both become negative and the potential becomes
unbounded from below.  The effects which differentiate between
their running, such as differing hypercharge assignments,
are small, and  a poor replacement for the
usual $h_t\gg h_b$ condition.  Furthermore, an $O(1)$ splitting between
$h_t$ and $h_{b,\tau}$ here is still of little use since it is
quickly reduced in importance by the proximity to  the fixed point.

Secondly, even when the electroweak symmetry is broken, a large
hierarchy of VEVs must be generated between the two similarly
evolving Higgs fields. To see the implications of this, let us
recall that the minimization of the tree level potential yields
\begin{equation}\label{sin2beta}
\sin 2 \beta={-2\mu_3^2\over \mu_1^2+\mu_2^2}.
\end{equation}
From here we see that large $\tan\beta$ will require $\mu_3^2$
to be small in magnitude relative to $\mu_1^2+\mu_2^2$.  
The second minimization
condition is
\begin{equation}
\tan^2\beta=\frac{\mu_1^2 + m_Z^2/2}{\mu_2^2 + m_Z^2/2},
\label{tanbeta1}
\end{equation}
or, equivalently,
\begin{equation}
m_Z^2={2(\mu_1^2-\mu_2^2 \tan^2\beta)\over \tan^2\beta-1}=
{2(m_{H_1}^2-m_{H_2}^2\tan^2\beta)\over \tan^2\beta-1}-2\mu^2,
\end{equation}
which in turn implies that for $\tan\beta\gg 1$ the solution we are
looking for requires $\mu_2^2\simeq -m_Z^2/2$, thus  setting the scale.
We, thus, see that a large hierarchy of VEVs requires the large
hierarchy $\mu_3^2\ll \mu_1^2+\mu_2^2$.  This, as is well known,
implies a degree of fine tuning between some of the parameters in the
Lagrangian.
Knowledge of $m_{H_1}^2$ and $m_{H_2}^2$ at $Q_0$ determines
$\mu^2(Q_0)$ from the approximate relation
\begin{equation}
\mu^2(Q_0)\simeq -m_{H_2}^2-{m_Z^2\over 2}+{m_{H_1}^2\over \tan^2\beta}
\end{equation}
and then  $B={\mu_3^2\over \mu}$ is also determined.  Note that
$|\mu_3^2|$ has to be small at large values of $\tan\beta$,
so that $B$ must also be small.  In numerical studies it has been
found that electroweak symmetry works well for reasonably large
$h_t$ so long as $h_b$ is neglected.  Upon its inclusion, the realization
of electroweak symmetry breaking engenders find tuning.  The dependence
of $\mu$ is somewhat complicated, although it may seem from
eq.(\ref{sin2beta}) that small $\mu$ gives large $\tan\beta$.
On the other hand $A$ (the trilinear soft SUSY breaking parameter)
seems to be relatively less important in absolute terms (i.e.,
one does not obtain any significant restriction on its magnitude), 
although it does influence the running of the soft-parameters.
In the NMSSM on the other hand, it plays an important role as we shall
see in the next section.  We note here that the MSSM spectrum is
unchanged if $M_{1/2}\rightarrow -M_{1/2}$ 
(where $M_{1/2}$ is the universal gaugino mass)
when $A=0$.

The main features of the sparticle spectrum turn out to be 
governed by (a) strong $\mu-M_{1/2}$ correlation, (b) large
values of $M_{1/2} \stackrel{>}{\sim}400\ $ GeV, and (c) $M_{1/2}
\stackrel{>}{\sim}
m_0$ (universal soft SUSY breaking scalar mass).  This implies small mixing
in the chargino and neutralino sectors.  The lightest supersymmetric
particle is mainly a bino, with a mass $\simeq 0.4\ M_{1/2}$.
The second lightest neutralino and lightest chargino are winos
and hence almost degenerate in mass.  The heaviest neutralino
and chargino are Dirac (pseudo-Dirac in the case of neutralino)
particles, with masses approximately equal to the parameter $|\mu(Q_0)|$.
An important tree level relation is obtained for large $\tan\beta>10$,
for which the tree-level value of the lightest CP-even higgs mass
($m_h$) is equal to $m_Z$, whenever the CP-odd boson mass ($m_A$)
is larger than $m_Z$, while for $m_A\leq m_Z, \ m_h=m_A$.
This tree level relation is approximately stable under radiative
corrections, with the only difference that the range for which
$m_h=m_A$ holds, extends to values of $m_A$ somewhat larger than
$m_Z$.  Therefore, large values of $\tan\beta$ and values of
CP-odd Higgs in the desired range imply $m_h<m_Z$.
We note that large values of $\tan\beta$ and values CP-odd higgs
mass $m_A<70\ $ GeV are preferred to improve the agreement with
the value of $R_b\equiv{\Gamma(Z\rightarrow b\overline{b})\over
\Gamma(Z\rightarrow {\rm hadrons})}$ measured at LEP~\cite{Antilogus}.

\setcounter{equation}{0}
\section{NMSSM at Large $\tan\beta$}

The potential for the Higgs fields of the NMSSM can be
obtained from eq.(\ref{superpotential}) and eq.(\ref{softterms})
through a standard procedure~\cite{Ellis}.  The minimization conditions
(evaluated at $Q_0$ after all the parameters are evolved via their
one-loop RG equations down to this scale), that determine the
soft SUSY breaking Higgs masses in terms of the other parameters, are

\begin{eqnarray}
m_{H_1}^2=-\lambda{{v_2}\over{v_1}} s (A_\lambda+ks) -\lambda^2(v_2^2+s^2)
+{{1}\over{4}}(g^2+g'^2)(v_2^2-v_1^2), \label{min1}\\
m_{H_2}^2=-\lambda {{v_1}\over{v_2}} s (A_\lambda+ks)-\lambda^2(v_1^2+s^2)
+{{1}\over{4}}(g^2+g'^2)(v_1^2-v_2^2),\label{min2}\\
m_{S}^2=-\lambda^2(v_1^2+v_2^2)-2k^2s^2-2\lambda s v_1 v_2 - kA_k s -
{{\lambda A_\lambda v_1 v_2}\over {s}}. \label{min3}
\end{eqnarray}
The first two minimization conditions can be rewritten as:
\begin{eqnarray}
\tan^2\beta={{m_Z^2/2+m_{H_1}^2+\lambda^2s^2}\over
{m_Z^2/2+m_{H_2}^2+\lambda^2s^2}},\label{tanbeta}  \\
\sin 2\beta={{(-2\lambda s)(A_\lambda+k s)}\over{m_{H_1}^2
+m_{H_2}^2+\lambda^2(2s^2+v^2)}}.\label{nsin2beta}
\end{eqnarray}
[Our normalization is such that 
$v^2\equiv v_1^2+v_2^2(=174 {\rm GeV})^2$ and $m_Z^2=\frac{1}{2}
(g^2+g'^2)v^2$, where $g$ and $g'$ are the gauge couplings of $SU(2)$
and $U(1)$, respectivly.]  These two equations give us some insight
into the manner in which our solutions are likely to behave.
Eq. (\ref{tanbeta}) guarantees that 
$\tan\beta$ must lie between 1 and $m_t/m_b$.
The proof, as in the case of the MSSM, relies once more on the
RG equations that govern the behaviour of mass parameters and may be
proved simply by reductio ad absurdum.
For this
purpose we need only consider the following equation expressing
the momentum dependence of the  difference of two supersymmetry
breaking scalar mass  parameters:
\begin{equation}
{{d}\over{dt}}(m_{H_1}^2-m_{H_2}^2)={{1}\over{8\pi^2}}
(-3h_t^2X_t+3h_b^2X_b+h_\tau^2X_\tau)
\end{equation}
\noindent where $t=\log(\mu)$, the logarithm of the
momentum scale, and $X_i$, $i=t,\ b,\ \tau$, are
combinations of scalar masses and trilinear couplings:
\begin{eqnarray}
X_t=m_{\tilde{Q}}^2+m_{\tilde{t}}^2+m_{H_2}^2+A_t^2, \nonumber \\
X_b=m_{\tilde{Q}}^2+m_{\tilde{b}}^2+m_{H_1}^2+A_b^2,  \\
X_\tau=m_{\tilde{L}}^2+m_{\tilde{\tau}}^2+m_{H_1}^2+A_\tau^2. \nonumber
\end{eqnarray}
It must be
noted that in order to prove that $\tan\beta > 1$ we neglect
$h_b$ and $h_\tau$, and for proving $\tan\beta < m_t/m_b$
we retain them.

From eq.(\ref{tanbeta}) it is clear that, as in MSSM, in order
to have large $\tan\beta$ with $m_{H_1}^2$ and $m_{H_2}^2$ being
essentially degenerate, because the top and bottom Yukawa couplings
are comparable in the RG evolution, the denominator of the equation has
to be small at the weak scale.  This implies the fine tuning
condition
\begin{equation}\label{finetune1}
m_{H_2}^2+\lambda^2s^2 \approx -m_Z^2/2.
\end{equation}
From this condition it follows that the correspondence with the MSSM
will occur in a certain well defined manner with the
identification of 
$\lambda s$ with $\mu$.  Similarly, we must identify
$A_\lambda+ks$ with $B$, the bilinear soft supersymmetry breaking
parameter of the MSSM.  We will show below that for large
values of $\tan\beta$ this identification occurs in a novel manner,
not generic to the model, say, for $\tan\beta \simeq 1$.
For large values of $\tan\beta$, eq.(\ref{nsin2beta}) implies
\begin{equation}
A_\lambda\approx -ks
\end{equation}
This is analogous to the condition $B\approx 0$ of the MSSM.  The
situation here is complicated because $A_\lambda$ is not a parameter
that is fixed at $Q_0$ but is present from the outset.  This is the
first of the fine tuning problems we encounter in the NMSSM.

A rearrangement of eq.(\ref{nsin2beta}) equation yields
\begin{equation}
\lambda s (A_\lambda+ks)=\tan\beta
(-m_{H_2}^2-\lambda^2s^2){{m_Z^2}\over{2}}{{(\tan\beta^2-1)}
\over{(\tan\beta^2+1)}}-{{\lambda^2v^2\sin 2\beta}
\over{2}}.
\end{equation}
In this equation we can discard the last term when $\tan\beta$
is large.  Then, with the identification of the appropriate parameters
in terms of those of the MSSM as described earlier, we recover all the
MSSM relations of the previous sections {\it without having to
go through a limiting procedure} that is required in the general
case.
The third minimization condition, eq.(\ref{min3}), can be rewritten
as:
\begin{equation}
m_S^2=-\lambda^2 v^2-2 A_\lambda^2+{{\lambda v^2 \sin 2\beta A_\lambda}
\over{k}}+A_\lambda A_k +{{\lambda k v^2 \sin 2\beta}\over{2}}.
\end{equation}
In order to satisfy this condition, one must have large
cancellations between the fourth and the first two terms,
since the terms proportional to $\sin 2\beta$ are negligible.  This
requires that $A_\lambda$ and $A_k$ come out with the same sign
and that their product be sufficiently large.  It is this
fine tuning condition that leads to problems with finding solutions
with sufficiently small trilinear couplings in magnitude~\cite{AP}.

The starting point of our analysis of the NMSSM is the 
 estimation of the scale
$M_X$ with the choice of the SUSY breaking scale $Q_0\sim 1\ TeV$.
For $\alpha_S(m_Z)=0.12$, $Q_0 = 1\ TeV$ and $\alpha=1/128$,
we find upon integrating the one-loop beta functions, $M_X=1.9\times
10^{16}\ GeV$, and the unified gauge coupling $\alpha_G(M_X)=1/25.6$.
We then choose a value for the unified Yukawa coupling $h$ of $O(1)$.
The free parameters of the model are the common gaugino mass $(M_{1/2})$,
the common scalar mass $(m_0)$, the common trilinear scalar coupling
$(A)$, and the two additional Yukawa coupligs $(\lambda,\ k)$, respectively.
Note that our convention requires us to choose $\lambda>0$ and
$k<0$ in order to conserve CP in the Yukawa sector of the model~\cite{AP}.
We also impose the
constraint that $|A|<3m_0$~ \cite{GRZ} in order to guarantee the absence of
electric-charge
breaking vacua.
  In the case at hand this choice may have to
be strengthened further due to the presence of large Yukawa couplings
for the b-quark.  The situation is considerably less restrictive
when mild non-universality is allowed and, for instance, if strict Yukawa
unification is relaxed.  Given these uncertainties, we choose to work
with this constraint.
 We also compute the mass of the charged Higgs boson~ \cite{Ellis}
\begin{equation}
m_C^2=m_W^2-\lambda^2v^2-\lambda(A_\lambda+ks){{2s}\over{\sin 2\beta}},
\end{equation}
\noindent where $m_W$ is the mass of the W-boson. We note that the
radiative corrections to the charged Higgs mass are small for most of
the parameter range, as in the case of MSSM~ \cite{Brignole}. The reason for this
is that a global $SU(2)\times SU(2)$ symmetry~ \cite{HP} protects the charged
Higgs mass from obtaining large radiative corrections.  If this quantity
were to come out to be negative, then the resulting vacuum would
break electric-charge spontaneously and the corresponding point
in the parameter
space would be excluded.

Choosing a particular set of values
for the input parameters $(M_{1/2},\ m_0,\ A,\ \lambda,\ k)$,
satisfying the above constraints, we calculate the crucial
parameters $\tan\beta$, $r(\equiv s/v)$, $A_\lambda$ and $A_k$
and $ks$, that determine the physical spectrum, from their
renormalization group evolution to $Q_0$.  We choose these
input parameters in such a manner so as to ensure that we are
in the neighborhood of a vacuum that breaks $SU(2)\times U(1)$~\cite{AP}.
With the boundary condition (1.1) at the unification scale, we
find $|M_{1/2}|\sim 500$ GeV for a wide range of values of
$\lambda \sim 0.4-0.6$ with $k=-0.1$, leading  to values of $\tan\beta\sim
60$.   
These magnitudes emerge when we choose $\lambda$ in a manner that does
not lead to very small values of $k$, and furthermore, we note that the 
solutions depend only on the ratio $\lambda/k$~\cite{AP}.
Once the sign of $k$ is fixed [from requirements
of CP invariance], a solution is found only when the sign of $M_{1/2}$
is negative.  Phenomenological requirements enforce $m_0$ also to
be rather large in order to guarantee that the lighter scalar tau
be heavier than the lightest neutralino. The minimization condition 
(3.3) or its equivalent (3.11)
enforces a large ratio $|A/m_0|\sim 3$.  It then follows
from eq.(3.9) that $s$ must be rather large in comparison with $m_Z$,
the only other scale in the problem.  We note that here we are 
near the fixed point of $\lambda$, and as such we expect that this provides
us with  a lower bound on the singlet vacuum expectation 
value $s$. Essentially, this implies that in order
to have Yukawa unification (1.1) in the NMSSM, and the resulting
large value of $\tan\beta$, the singlet vacuum expectation value
must be large compared to the doublet vacuum expectation values.
As we shall see in Sec. 4, these conclusions are borne out by the
detailed numerical calculations.

We now analyze the implications of the large values of the singlet
vacuum expectation value $s$, that arises in the NMSSM with eq. (1.1),
on the mass spectra.  In the basis $({\rm Im} H_1^0, {\rm Im} H_2^0,
{\rm Im} S)$, with neutral Goldstone boson 
$G^0=\cos\beta ({\rm Im} H_1^0)-\sin\beta ({\rm Im} H_2^0)$
and the orthogonal combination
$A^0=\cos\beta ({\rm Im} H_1^0)+\sin\beta ({\rm Im} H_2^0)$,
the tree-level mass squared matrix for the two pseudoscalar neutral Higgs
fields can be written as (in numerical calculations we take into account
one-loop radiative corrections in the effective potential approximation)
\begin{equation}
M_P^2=\left(\begin{array}{c c}
{2\lambda s A_\Sigma \over \sin 2\beta} & \lambda v (A_\Sigma+3ks) \\
\lambda v (A_\Sigma+3ks) & -3ksA_k +{\lambda v^2 \sin 2\beta \over 2s}
(A_\Sigma -3ks) \\
\end{array} \right),
\end{equation}
where
$A_\Sigma=-(A_\lambda+ks)$.  The pseudoscalar masses are
\begin{equation}
m_{P_1,P_2}^2={1\over 2}[M_{22}+M_{11}\mp \sqrt{(M_{22}-M_{11})^2+4 M_{12}^2}],
\end{equation}
where
\begin{eqnarray}
M_{22}\pm M_{11}={\lambda v^2 \sin 2\beta \over 2s}(A_{\Sigma}-3ks)
-3ksA_k\pm{2\lambda s A_\Sigma \over \sin 2\beta}, \nonumber \\
M_{12}=\lambda v (A_{\Sigma}+3ks).
\end{eqnarray}
The pseudoscalar eigenstates are described through the mixing angle $\gamma$
given by:
\begin{equation}
\sin 2\gamma={-2M_{12}\over \sqrt{(M_{22}-M_{11})^2+4 M_{12}^2}}.
\end{equation}
For large values of $s$, the pseudoscalar Higgs masses are approximately given 
by
\begin{equation}
m_{P_1}^2\simeq -3ksA_k,\ m_{P_2}^2\simeq {2\lambda s A_\Sigma\over
\sin 2\beta}.
\end{equation}

This shows that the pseudoscalar $P_1$ is mainly ${\rm Im} S$, whereas
$P_2$ is a mixture of ${\rm Im}H_1^0$ and ${\rm Im}H_2^0$.
Furthermore, for large values of $\tan\beta$, $P_2$ becomes very
massive.  Thus $P_1$ decouples from the spectrum because it is
mainly a singlet and $P_2$ decouples from the spectrum because
it is very massive simplying that the pseudoscalar Higgs bosons
would be impossible to produce in the limit of Yukawa unification.
These qualitative features of the pseudoscalar spectrum 
survive the effects of radiative corrections.
Here {\it we find a qualitative distinction  with the corresponding
pseudoscalar spectrum of the MSSM.}
On the other hand the tree level the squared mass matrix,
$M_S^2$, for the neutral scalar Higgs bosons
can be written as
\begin{small}
\begin{footnotesize}
\begin{equation}
\left( \begin{array}{c c c}
m_Z^2\cos^2\beta+A_\Sigma s \tan\beta & -\lambda s A_\Sigma+
\lambda^2 v^2 \sin 2\beta -{m_Z^2\sin 2\beta \over 2} &
\lambda v_2 (2\lambda s \cot\beta + ks -A_\Sigma) \\
-\lambda s A_\Sigma+
\lambda^2 v^2 \sin 2\beta -{m_Z^2\sin 2\beta \over 2} &
m_Z^2\sin^2\beta + A_\Sigma s \cot\beta & \lambda v_1(2\lambda s \tan\beta
+ks -A_\Sigma) \\
\lambda v_2 (2\lambda s \cot\beta + ks -A_\Sigma) & 
\lambda v_1(2\lambda s \tan\beta
+ks -A_\Sigma)
  & (4k^2s^2+ksa_k)-{\lambda A_\lambda v^2 \sin 2\beta \over 2s}\\
\end{array} \right).
\end{equation}
\end{footnotesize}
\end{small}
We shall label the CP-even Higgs eigenstates of eq.(3.18) as
$S_1,\ S_2$, and  $S_3$, respectively in order of increasing mass.
It is not very illuminating to present complicated analytical expressions
for them, and so  we will defer the numerical results to
Sec. 4.  It turns out that in the limit of large
$\tan\beta$, the lightest CP-even Higgs boson is almost a pure
${\rm Re} (H^0_2)$, whereas the heaviest CP-even Higgs boson is
predominantly the  singlet, ${\rm Re} S$.  Furthermore, the lightest
Higgs boson mass has an upper bound which is close to the upper
bound in the MSSM in the same limit.  This is a consequence of
the fact that the upper bound on the lightest Higgs mass in the two 
models, including radiative corrections,  differs by a term that is 
proportional to $\sin 2\beta$,
and is, 
therefore, small~ \cite{PNP1}.

The neutralino mass matrix ($M_{\chi^0}$) in the Non Minimal Supersymmetric 
Standard model can be written as~ \cite{PNP2} :
\begin{footnotesize}
\begin{equation}\label{neutralino}
\left( \begin{array}{c c c c c}
M_1 & 0 & -m_Z\cos\beta \sin \theta_w & m_Z\sin\beta\sin\theta_W & 0 \\
0 & M_2 & m_Z\cos\beta \cos\theta_W & -m_Z\sin\beta\cos\theta_w & 0 \\
 -m_Z\cos\beta \sin \theta_w & m_Z\cos\beta \cos\theta_W & 0 & \lambda s &
\lambda v \sin\beta \\
m_Z\sin\beta\sin\theta_W & -m_Z\sin\beta\cos\theta_w & \lambda s & 0 &
\lambda v \cos\beta \\
0 & 0 & \lambda v \sin\beta & \lambda v \cos\beta & 2ks \\
\end{array} \right),
\end{equation}
\end{footnotesize}
where $M_1$ and $M_2$ are  the masses of the $U(1)_Y$ and $SU(2)_L$
gauginos ($\lambda^{\prime}; \lambda^a, a = 1, 2, 3$), respectively, and where we have chosen the basis
$(-i\lambda^{\prime},-i\lambda^3, \Psi_{H_1^1}, \Psi_{H_2^2}, \Psi_S)$.
The grand unification condition leads to the mass relation
$M_1=(5/3)M_2\tan^2\theta_W$.  The gluino mass is $M_3$; the gluino
does not mix with the rest of the neutralinos.  Because neutralinos
are Majorana particles, the mass matrix (3.19) is in general comples 
symmetric, and hence can be diagonalized by only one unitary matrix
N, i.e., $N^{\ast} M_{\chi^0} N^{-1} = M_{\chi^0}^D$. We shall label the 
neutralino masses in the ascending order (of magnitude) 
as $M_{\chi^0_i}$, 
i = 1,.....,5. 

In the limit of large $s$, we have the 
the approximate expressions for the (magnitude of) 
 neutralino masses in increasing  order:
\begin{equation}
M_1,\ M_2,\ \lambda s, \lambda s, 2ks,
\end{equation}
with the lightest neutralino being a predominantly a bino, which satisfies
the simple mass relation
\begin{equation}
M_1\simeq {\alpha_1(Q_0) \over \alpha_G}M_{1/2} \simeq 0.45 M_{1/2}.
\end{equation}
All the neutralinos with the exception of the heaviest one have
negligible singlet component;  the singlet decouples from the remainder
of the spectrum. We note that two of the neutralinos are nearly degenerate,
thus leading to a pseudo-Dirac neutralino, as in MSSM, in the limit of 
large $\tan\beta$.

The chargino mass matrix $M_{\chi^\pm}$ is given by:
\begin{equation}\label{chargino}
\left( \begin{array}{ c c}
M_2 & \sqrt{2} m_W \sin\beta \\
\sqrt{2} m_W\cos\beta & -\lambda s
\end{array} \right)
\end{equation}
and is same as in  MSSM with the role of $\mu$ played by $\lambda s$.
The masses and the composition of the charginos
are obtained by diagonalizing the matrix (3.22) 
via the biunitary transformation
$U M_{\chi^\pm} V^{-1} = M_{\chi^\pm}^D$, where U and V are $2 \times 2$ 
matrices which diagonalize the hermetian matrices 
$M_{\chi^\pm} M_{\chi^{\pm}}^{\dagger}$ and 
$M_{\chi^\pm}^{\dagger} M_{\chi^\pm}$,
respectively. Similarly, the masses of other sparticles are same as in MSSM
with $\mu$ replaced by $\lambda s$.

\setcounter{equation}{0}
\section{Numerical Results and Discussion}


Having described the NMSSM with large $\tan\beta$ in detail in the
previous section, we now turn to obtaining the  particle spectrum of the model
numerically.

Writing down the coupled set of the RG equations for the 24 parameters
of the model~\cite{DS,NKF}, and including the contributions
of $h_b$ and $h_\tau$~\cite{AP}, the parameter space of the model is
scanned by taking values of the input parameters
$(M_{1/2}, m_0, A, h,\lambda,k)$ at $M_X$ which are evolved down
to low  energies $Q_0$ to obtain the values of
of the parameters $\tan\beta$, $r\equiv s/v$, $A_\lambda$,
$A_k$, $ks$ .  We also evolve  the
values of the soft masses appearing on the left
hand sides of eqs. (\ref{min1})  -- (\ref{min3}),
and compare their values to the combinations of the parameters
appearing on the right hand sides of these equations as  obtained
from the RG evolution.  The input parameters are chosen so
as to satisfy the constraints described in the previous section,
namely the absence of electric charge breaking vacua, the
charged Higgs mass squared remaining positive, and the $SU(2)\times
U(1)$ breaking minimum being energetically favorable.  It turns out
that the first of the minimization conditions, eq.(\ref{min1}),
 is the one which is
most sensitive to the choice of initial conditions reflecting
the fine tuning discussed in Sec. 3.  The parameters are chosen
so as to study
$r_1$, $r_2$ and $r_3$, which are defined as
the difference between the left and right hand sides of the
three minimization eqs. (3.1) -- (3.2) divided by the
right hand side of
each of these equations, and study the change in sign that
these suffer as the parameters are varied.
There are enormous difficulties in trying to achieve a simultaneous
solution to $r_i=0,\ i=1,2,3$.  In particular $r_3=0$ requires the
presence of values for $|A|/m_0$ of almost 3 or more.  This
requirement has a profound impact on the particle spectrum of
the model.  In particular, $r$, the ratio of the singlet to the doublet
vacuum expectation value persistently remains large for the choice
of parameters considered with large $\tan\beta$, corresponding to
the VEV of $s$ being of the order of $10$ TeV.  This is substantially
different from the situation when $\tan\beta\simeq 1$~\cite{Ellis}.
We note that the singlet vacuum expectation value is not
constrained by the experimental data.

In Fig. 1, we show a typical evolution of the three soft SUSY
breaking mass parameters $m_{H_1}^2$,  $m_{H_2}^2$ and  $m_{S}^2$
from $M_X$ down to the low scale $Q_0$ with a choice of parameters
such that all constraints are satisfied, and we are in the neighbourhood
of an $SU(2)\times U(1)$ breaking vacuum.  We note that because of
the possibility of large value of $m_t(m_t)=181$ GeV, we have
a large value for $h$ so that the Yukawa couplings dominate  over the
gauge couplings in the 
evolution of these parameters.  This in turn
forces the mass parameters to remain large at large momentum scales
compared to their values at smaller momentum scales.

In supersymmetric theories with R-parity conservation, the lightest
supersymmetric particle generally turns out to be the lightest neutralino.
In the NMSSM the neutralino mass matrix is given by
eq.(\ref{neutralino}) and its general properties 
are discussed in~\cite{PNP2}.
The parameters that determine the mass matrix are $\lambda, k,
s,\tan\beta,M_1$ and $M_2$.  Choosing the input parameters at $M_X$ so
that all the constraints of the Sec. 3 are satisfied, we obtain
values for these parameters at $Q_0$ and then the neutralino mass
matrix may be evaluated numerically.  The chargino mass matrix
eq.(\ref{chargino}) may also be evaluated in a  similar manner.  

One result of this procedure is shown in Fig. 2, where we plot the
lightest neutralino and chargino masses for a specific 
choice of input parameters
of Table 1, as obtained from RG evolution, 
as a function of the top quark mass.
We have found from our scan of the parameter space, 
that the lightest neutralino is
almost a pure bino in the limit of large $\tan\beta$.  Furthermore,
all other neutralinos except the heaviest one have a negligible
singlet component, indicating that the singlet completely
decouples from the lighter neutralino spectrum.  These properties 
of the neutralino spectrum are shown in the second and 
third columns of Table 2. The mass of the
lightest  neutralino in this case
is determined by the simple mass relation for the bino
$
M_1\simeq \left( {\alpha_1(M_G)\over \alpha_G}\right) M_{1/2}
\simeq 0.45 M_{1/2}.
$
The masses of the heavier neutralinos lie in the range of $0.5-1$ TeV.
Furthermore, the lightest chargino mass
bears a relation to $M_{1/2}$ similar to the neutralino relation,
with $\alpha_1$ in eq. (3.21) replaced by $\alpha_2$, reflecting that
it is primarily a charged wino. This is seen from the fourth, fifth and 
sixth columns of Table 2.  The heavier chargino mass is found to
be $\simeq 1$ TeV. We further note that two of the neutralinos are 
nearly degenerate, and lead to a pseudo-Dirac neutralino. Also the second
lightest neutralino is primarily a wino and degenerate in mass with the 
lightest chargino. These characteristics are similar to those found 
in MSSM.  The gluino mass is found to be $1.6$ TeV for
the choice of parameters of Fig. 2 and follows from a relation
similar to eq. (12) with  $\alpha_1$ replaced by  $\alpha_S$.

We  now come to the spectrum of CP-even Higgs bosons of the model.
In  order to understand the quantitative features
of the results we have obtained for the lightest CP-even Higgs boson,
we need to go into some detail regarding the actual choices of parameters
entering the computation and the correlations between the various elements
of the spectrum. 
We have already reviewed the corresponding situation in the case of
the MSSM in Sec. 2,  and here we will discuss the comparison of the
NMSSM with that of the MSSM.
In Fig. 3, we plot for typical and reasonable values of the input
parameters, in the region where the vacuum is expected to lie, the
mass of the lightest CP-even Higgs boson as a function of the
top-quark mass $m_t(m_t)$ in the range that is most favoured under
these boundary conditions~\cite{ALS1,ALS2,OP,Anderson,AK}.  
The choice of parameters here is closely related to the family
of solutions studied extensively in Ref.~\cite{AP},  and would serve
as a typical example of the numbers we have explored. 
In the MSSM when the mass of its unique CP-odd Higgs 
boson $m_A\gg m_Z$, the substantive part of the radiative correction
is picked up by the lighter of the CP-even bosons, $h^0$.  
As $m_A$ approaches
$m_Z$, the radiative corrections are  shifted to the heavier
of the CP-even Higgs bosons, $H^0$.  Such a feature is observed here
as well:
for those choices of parameters in Table 1 that yield a somewhat smaller
$m_{P_1}$ (the mass of lightest CP-odd Higgs boson in NMSSM), 
we find that the radiative corrections to the lighest
CP-even Higgs, $S_1$, are smaller.
We note, however,  that the CP-odd Higgs bosons here are always
massive in the limit of large $\tan\beta$ in contrast to the
situation that prevails in the MSSM, where $m_A$ even in the
vicinity of $m_Z$ is plausible~\cite{ALS2}.
  Due to the complexity of the system
under investigation and the difficulty in controlling the
numerical choice of the parameter $\lambda$ for
a given $h$,  with all other parameters held fixed, we do not know
how precisely close the choice of parameters of Table 1 are to
a genuine ground state. 
Furthermore, we note that
the clarity with which the correlations have been
observed  between $m_A$ and $m_{h^0}$ in MSSM do not have
a simple parallel here due to the presence of a larger  number of
physical states.
A more precise, albeit prohibitively time consuming, determination could then
ensure that the spurious wobble seen in Fig. 3 is eliminated,
and would establish a more reliable correlation between increasing
$h$ and the rise of the mass of $S_1$ and the correlations
with $m_{P_1}$.
Furthermore, a refinement of the choice of parameters,
based on the minimization of the
one-loop effective potential,  could stablize the figures presented
here.

We note that the lightest
Higgs bosons mass $\sim 130$ GeV for a wide range of parameters
which nearly saturates the upper bound of $140$ GeV~\cite{AP}, 
and lies in the 
same range as in 
MSSM with large $\tan\beta$.  This is 
a consequence of the largeness of $\tan\beta$:
the contribution to the tree level
mass which depends on the trilinear coulings $\lambda$ is small,
being proportional to $\sin^2 2\beta$, so that the upper bound on
the lightest Higgs mass reduces to the corresponding upper bound 
in MSSM when the appropriate identification of the parameters
is performed~ \cite{PNP1}. 
We also note that the upper bound on the lightest Higgs mass depends
only logarithmically on $r$, and hence on the singlet vacuum
expectation value $s$, in the limit of large $r$, which, therefore
decouples from the bound~\cite{CV}.  Furthermore, the lightest Higgs
boson in almost a pure doublet Higgs (${\rm Re}\ H_2^0$), with
the singlet component being less that $1\%$ in the entire range
of parameters considered.  It is only the second heavier CP  even
Higgs boson $S_2$ that is predominantly a singlet.  Its mass ranges
between $740$ GeV and $2.3$ TeV.  The heaviest 
CP even Higgs boson $S_3$ is again
predominantly a doublet Higgs  (${\rm Re}\ H_1^0$) with its mass
varying between $4-6$ TeV.  This implies that all the CP-even
Higgs bosons, except the lightest one, decouple from the spectrum.
These features of the spectrum of the CP-even Higgs bosons of NMSSM 
for large values of $\tan\beta$ are clearly seen  from Table 3, where we 
show the mass and composition of $S_i$ for a wide range of input parameters.
The results presented above, that the lightest Higgs boson is
almost purely a doublet Higgs at large $\tan\beta$, are in contrast
to the situation with low values of $\tan\beta$, where the
lightest CP-even Higgs boson contains a large admixture of the
gauge singlet field $S$~\cite{Ellis,UE,TE}. 

In Fig. 4 we plot for the typical values of the input parameters of Table 1
the mass of the lightest CP-odd Higgs boson as a function of $m_t(m_t)$.
We note from Table 3 that both CP-odd Higgs bosons $P_1$ and $P_2$  are 
heavy, their masses being in the range
of $2$ TeV and $6$ TeV, respectively.  Also, the lightest CP-odd
state is predominatly a Higgs singlet, thereby effectively
decoupling from the rest of the spectrum.  The charged Higgs boson 
mass $m_C$ lies, for most of the cases that we have studied, in the range
$1-2$ TeV. 

In order to discuss the features of the 
the sfermion spectrum, we first recall some of the features of
the spectrum of the MSSM.
In the MSSM  
it has been observed~\cite{ALS2} that the presence of large
Yukawa couplings for the b-quark as well as the $\tau$ lepton,
as well as the presence of large trilinear couplings,
could lead to the lighter of the scalar $\tau$'s tending to become
lighter than the lightest neutralino, which is the most favoured
candidate in such models for the lightest supersymmetric particle.
In particular, in order to overcome cosmological constraints
for given values of $M_{1/2}$, lower bounds on $m_0$ were found
to emerge.  In turn, increasing $m_0$ implies ever decreasing
$m_A$ (clear correlations have been described for the case of
$A=0$ in Ref.~\cite{ALS2}) thus leading to further constraints
on the parameter space of the MSSM.
For large values of $\tan\beta$ the competing tendencies between the 
lighter scalar tau mass and $m_A$ 
play an important role in MSSM
in establishing a lower bound $\sim 450$ GeV on $M_{1/2}$. 
Given the 
complexity of the system of equations, it has not been possible to
to extract similar lower bounds on $M_{1/2}$ in NMSSM. 
Nevertheless, in Ref.~\cite{AP}
the intimate link between the ground states of the two models has been 
established and a much more sophisticated and time consuming analysis of 
the present model is also likely to yield a lower bound that is unlikely
to be very different from the one obtained in MSSM. As a result, in confining
ourselves to numbers of this magnitude and higher, we find a heavy spectrum. 
More recently~\cite{Matalliotakis} further experimental constraints
on MSSM have been taken into account resulting in an extension of the
minimal assumptions at $M_X$ by including non-universality for scalar
masses.  Indeed, in the present analysis similar problems
have been encountered with some of the choice of parameters studied
in Ref.~\cite{AP}, with $m_{\tilde{\tau}_1}$ tending to lie below
the mass of the lightest neutralino due to the persistent presence
of large Yukawa couplings and more so due to the large trilinear
couplings dictated by eq.(\ref{min3}).  Nevertheless, given the fact
that the present work minimizes the tree-level potential and that
the violations of cosmological constraints are not serious, in that
minor adjustments of $|A|/m_0$ solve this problem efficiently, we
consider the regions of the parameter space we have explored to
be reasonable.  Furthermore, it could be that the extension of
minimal boundary conditions along the lines of Ref.~\cite{Matalliotakis}
could provide alternative and elegant solutions to this problem,
while preserving the existence of relatively light scalar $\tau$'s
as a prediction of the unification of Yukawa couplings in the NMSSM
as well as in the MSSM.

The heaviest sfermions in the spectrum of NMSSM,  as in
MSSM, are the scalar quarks, which 
tend to be much heavier, in the TeV range.  The $SO(10)$ property
that the scalar b-quarks are as massive as the scalar top-quarks is
preserved in the NMSSM.

To summarize,
through a detailed analysis of NMSSM presented here, we have shown that
all the particles, except the lightest CP - even Higgs boson, implied by
supersymmetry are heavy for large values of $\tan\beta$.  The
gauge singlet field S decouples,  both from the lightest Higgs
boson as well as the neutralinos. 
The LSP of the model continues to be, as in MSSM, the lightest
neutralino that is primarily a bino in composition, with the lighter
scalar $\tau$ having a mass in the neighbourhood of the LSP mass.
The remainder of the spectrum tends to be heavy, from 1 to a few TeV.
We note that the NMSSM in the large $\tan\beta$ regime
rests on a delicately hinged system of
equations and constraints.  Although it provides a good testing
ground for the stability of the predictions of the MSSM, in
practice it deserves great care in its treatment.

\noindent{\bf Acknowledgements}:
The research of BA is  supported by the Swiss National
Science Foundation.
PNP thanks the Alexander von Humboldt-Stiftung
and Universit\"at Kaiserslautern, especially
Prof. H. J. W. M\"uller-Kirsten, for support 
while this work was completed.
The work of PNP is supported by the
Department of Science and
Technology, India under
Grant No. SP/S2/K-17/94.

\newpage

\bigskip

\noindent{\Large{\bf Table Captions}}

\bigskip

\noindent {\bf Table 1}  Sample of values of input parameters ($M_{1/2}$, 
$m_0$, $A$, $h$, $k$, $\lambda$) and the computed values of different parameters 
($ m_t(m_t)$, $\tan\beta$, $r$, $A_{\lambda}$, $A_k$  and $ks$). 
All mass parameters 
are in units of GeV.

\bigskip

\noindent {\bf Table 2} Masses and compositions of neutralino $\chi_i^0$ and 
chargino $\chi_i^{\pm}$ states for the values of input parameters  of Table 1.
All masses are in GeV.

\bigskip

\noindent {\bf Table 3} Masses and compositions of the CP-even ($S_i$) 
and CP-odd ($P_i$) Higgs bosons. The composition of CP-even Higgs 
bosons is in terms of the basis ($Re H_1^0,~ ReH_2^0,~ Re S$), whereas the 
basis for CP-odd Higgs bosons is ($A^0,~ Im S$). All masses are in GeV.

\bigskip

\bigskip 

\noindent{\Large{\bf Figure Captions}}

\bigskip

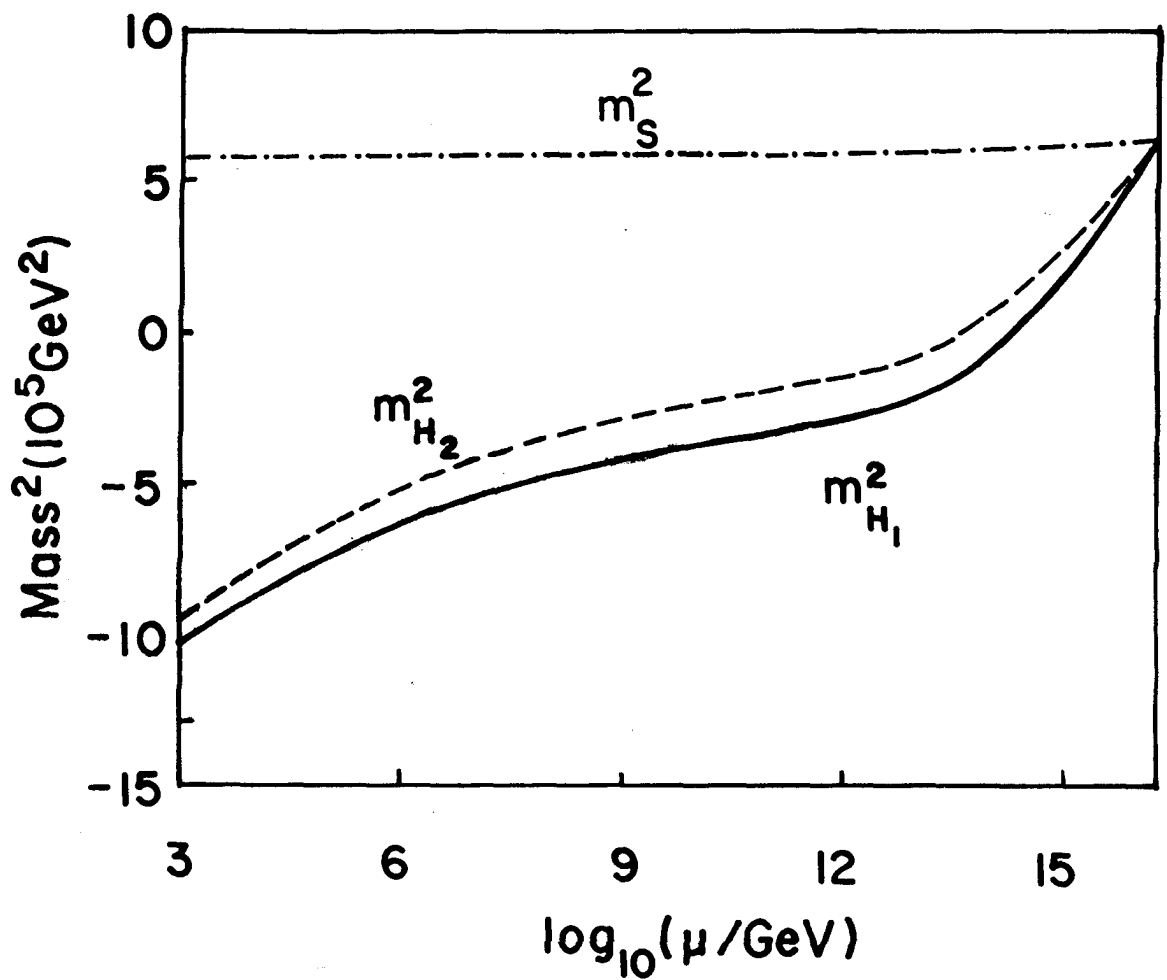
\noindent {\bf Fig. 1}  The evolution of soft supersymmetry breaking
mass parameters from the grand unified scale $M_X$ to $Q_0$ defined
in the text.  The input parameters are $M_{1/2}=-700$, $m_0$ and
$A=1600$ (all in GeV).  The other parameters are $h=1.5,\
\lambda=0.40$ and $k=-0.10$.  The associated value of the top
quark mass is $181$ GeV.

\bigskip

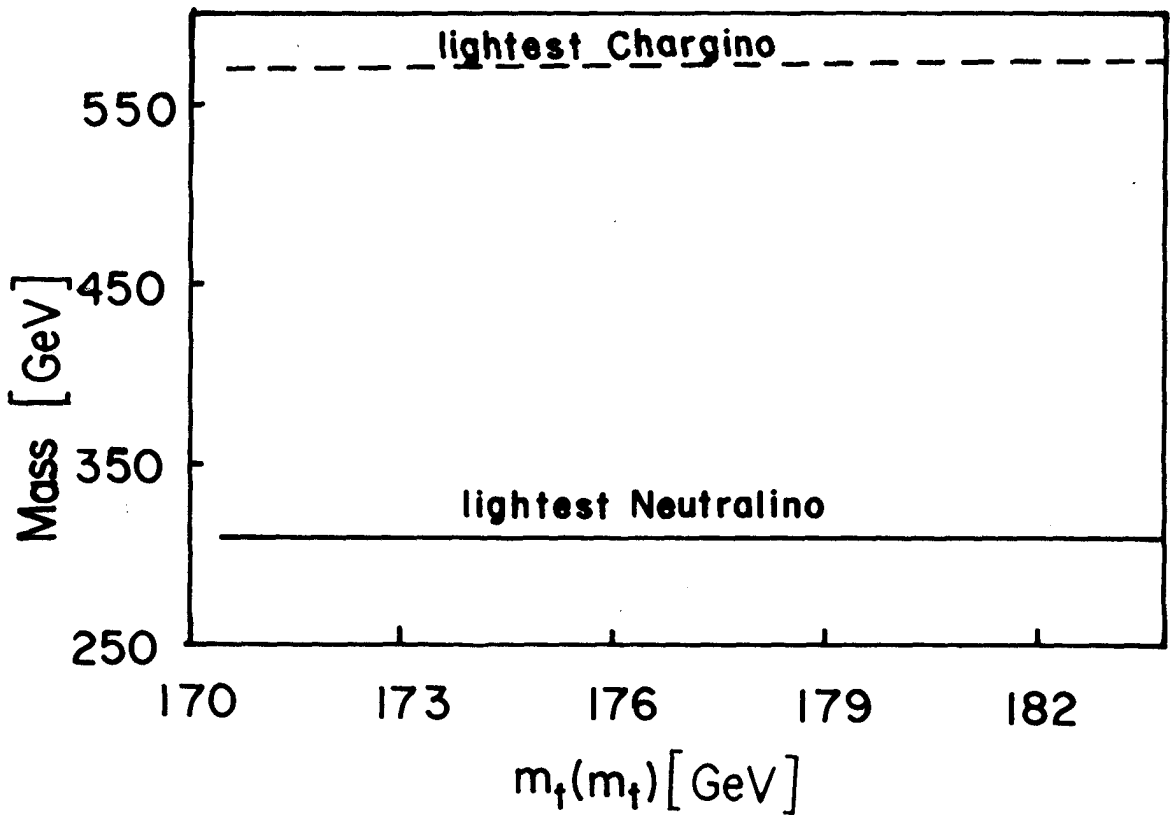
\noindent {\bf Fig. 2} The lightest neutralino and chargino masses 
as a function of $m_t(m_t)$.  The input parameters are $M_{1/2}=-700,
\ m_0=800,\ A=2200$ (all in GeV), with the remaining parameters
varied to guarantee a solution.

\bigskip

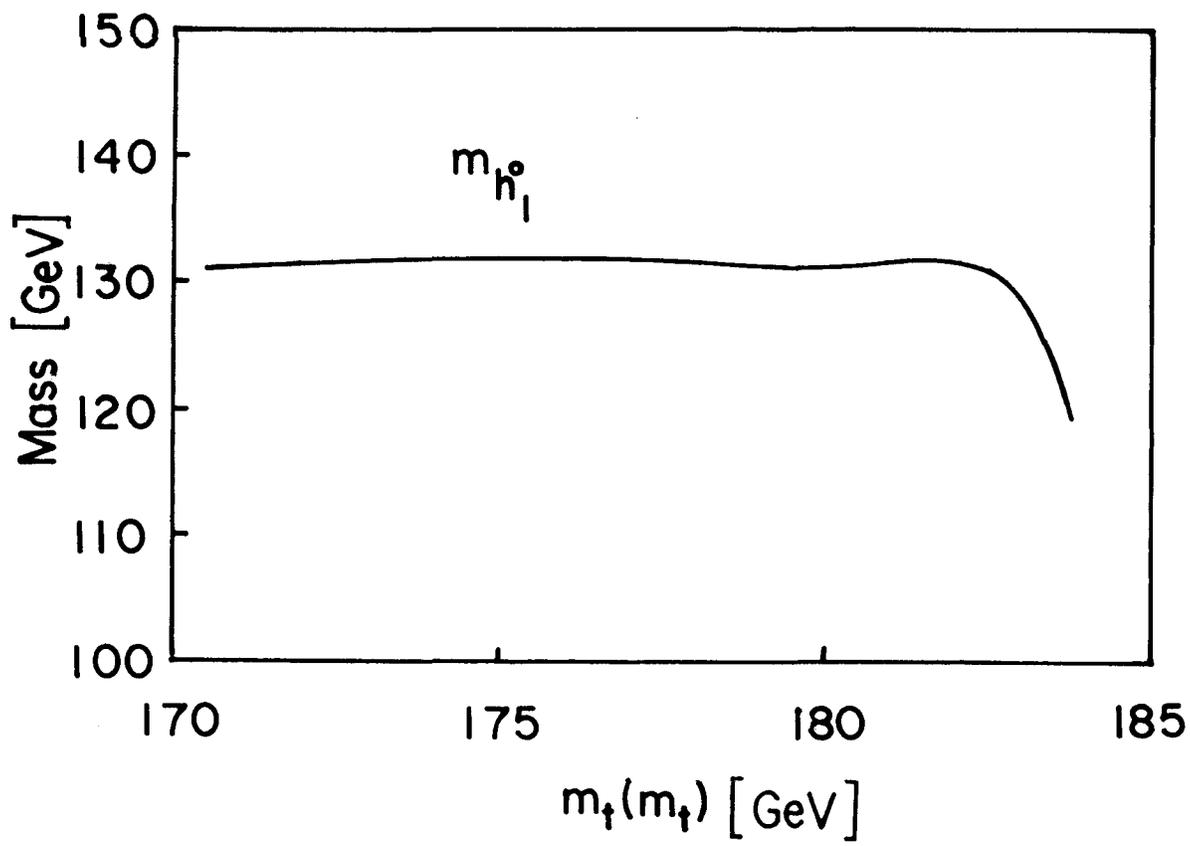
\noindent {\bf Fig. 3} The lightest CP-even Higgs boson mass as a function
of $m_t$.  The range of parameters is as in Fig. 2.

\bigskip

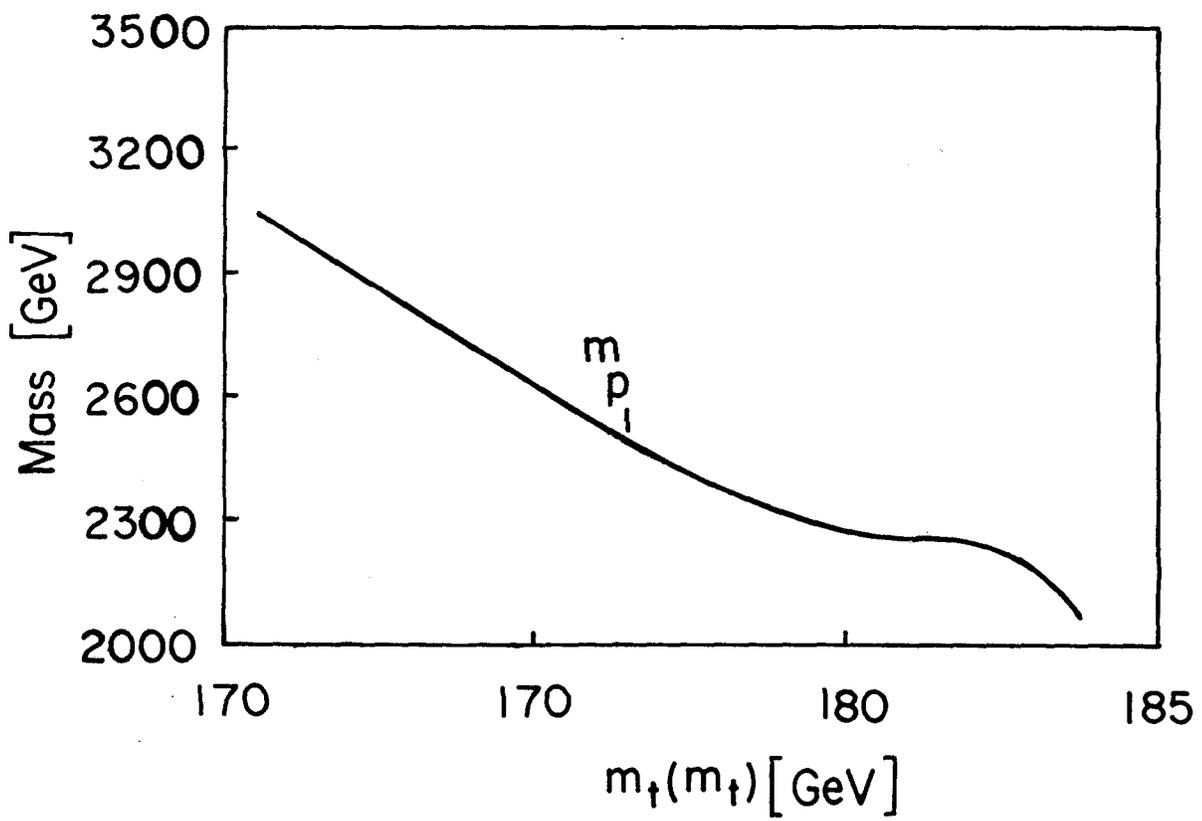
\noindent {\bf Fig. 4} The lightest CP-odd  Higgs boson mass as a function
of $m_t$. The range of parameters is as in Fig. 2.

\newpage

\vskip 5cm

\begin{small}
$$
\begin{array}
{||c|c|c|c|c|c|c||c|c|c|c|c|c||}\hline
\# & M_{1/2} & m_0 &  A & h & \lambda & k & m_t(m_t) & \tan\beta & r &  
A_{\lambda} & A_k & ks \\ \hline
1 & -700 & 800 & 2200 & 0.75 & 0.1 & -0.1 & 170.5 & 54 & 85 & 908 & 2131 & 
-1453 \\
2 & -700 & 800 & 2200 & 1.00 & 0.2 & -0.1 & 176.5 & 58 & 58 & 807 & 2100 &
-975 \\
3 & -700 & 800 & 2200 & 1.25 & 0.3 & -0.1 & 179.6 & 60 & 49 & 745 & 2070 &
-807 \\
4 & -700 & 800 & 2200 & 1.50 & 0.4 & -0.1 & 181.4 & 62 & 44 & 703 & 2045 &
-717 \\
5 & -700 & 800 & 2200 & 1.75 & 0.4 & -0.1 & 182.5 & 63 & 51 & 698 & 2067 &
-835 \\
6 & -700 & 800 & 2200 & 2.00 & 0.5 & -0.1 & 183.3 & 64 & 47 & 675 & 2048 &
-755 \\
7 & -700 & 800 & 2200 & 2.25 & 0.6 & -0.1 & 183.8 & 64 & 44 & 657 & 2032 &
-700 \\ \hline
\end{array}
$$
\end{small}
$$
{\rm Table \ 1}
$$

\newpage

\vspace*{-3.5cm}

\renewcommand{\baselinestretch}{1.35}

\begin{small}
$$
\begin{array}{|c|c|c|c|c|c|}\hline
\# & M_{\chi_i^0} & N_{ij} &M_{\chi_i^\pm} & V_{ij} & U_{ij} \\ \hline
1 & -309 & 0.99,-0.01,-0.06,0.02,0.00 & 579 & 0.99,0.15 & 0.97,0.22 \\
  & -579 & 0.02,0.98,0.16,-0.11,0.00 & & & \\
  & 877 & 0.03,-0.04,0.70,0.70,0.00 & & & \\
  & -887 & 0.05,-0.20,0.70,-0.70,0.00 & & & \\
  & -2906 & 0.00,0.00,0.00,0.00,0.99 & 886 &-0.15,0.99 & -0.22,0.97 \\ \hline
2 & -309 & 0.99,-0.01,-0.05,0.02,0.00 & 581 & 0.99,0.12 & 0.97,0.22 \\
  & -580 & 0.01,0.99,0.14,-0.09,0.00 & & & \\
  & 941 & 0.02,-0.04,0.70,0.70,0.00 & & & \\
  & -949 & 0.05,-0.16,0.70,-0.70,0.01 & & & \\
  & -1950 & 0.00,0.00,-0.01,0.00,0.99 & 949 &-0.12,0.99 & -0.22,0.97 \\ \hline
3 & -309 & 0.99,-0.06,-0.05,0.02,0.00 & 581 & 0.99,0.10 & 0.98,0.19 \\
  & -581 & 0.01,0.97,0.14,-0.09,0.00 & & & \\
  & 936 & 0.02,-0.04,0.70,0.70,0.00 & & & \\
  & -944 & 0.05,-0.16,0.69,-0.70,0.00 & & & \\
  & -1613 & 0.00,0.00,-0.02,0.01,0.99 & 945 &-0.10,0.99 & -0.19,0.98 \\ \hline
4 & -309 & 0.99,0.00,-0.05,0.02,0.00 & 582 & 0.99,0.12 & 0.99,0.07 \\
  & -582 & 0.01,0.99,0.12,-0.07,0.00 & & & \\
  & 1001 & 0.02,-0.03,0.70,0.70,0.00 & & & \\
  & -1007 & 0.04,-0.14,0.70,-0.7,0.04 & & & \\
  & -1436 & 0.00,0.00,-0.03,0.02,0.99 & 1008 &-0.12,0.99 & -0.07,0.99 \\ \hline
5 & -309 & 0.99,0.00,-0.05,0.02,0.00 & 582 & 0.99,0.10 & 0.99,0.17 \\
  & -582 & 0.01,0.99,0.13,-0.08,0.00 & & & \\
  & 979 & 0.02,-0.03,0.70,0.70,0.00 & & & \\
  & -986 & 0.05,-0.14,0.70,-0.7,0.00 & & & \\
  & -1669 & -0.33,0.00,-0.02,0.00,0.99 & 1018 &-0.10,0.99 & -0.17,0.98 \\ \hline
6 & -309 & 0.99,0.00,-0.05,0.02,0.00 & 581 & 0.99,0.10 & 0.94,0.35 \\
  & -582 & 0.01,0.99,0.13,-0.08,0.00 & & & \\
  & 978 & 0.02,-0.04,0.70,0.70,0.00 & & & \\
  & -984 & 0.05,-0.14,0.70,-0.7,0.00 & & & \\
  & -1509 & 0.00,0.00,-0.02,0.01,0.99 & 985 &-0.10,0.99 & -0.35,0.94 \\ \hline
7 & -309 & 0.99,0.00,-0.05,0.01,0.00 & 582 & 0.99,0.10 & 0.98,0.17 \\
  & -582 & 0.01,0.99,0.12,-0.07,0.00 & & & \\
  & 998 & 0.02,-0.03,0.70,0.70,0.00 & & & \\
  & -1004 & 0.05,-0.14,0.70,-0.7,0.04 & & & \\
  & -1402 & 0.00,0.00,-0.03,0.03,0.99 & 1004 &-0.10,0.99 & -0.17,0.98 \\ \hline
\end{array}
$$
\end{small}
\vspace{-4mm}
$$
{\rm Table \ 2}
$$

\newpage

\renewcommand{\baselinestretch}{1.7}
\begin{small}
$$
\begin{array}{|c||c|c||c|c|}\hline
\# & m_{S_i} & {\rm Composition} & m_{P_i} & {\rm Composition} \\ \hline
1 & 131 & -0.02,0.99,0.00 & 3048 & 0.00,1.00 \\
  & 2314 & 0.00,0.00,0.99 & & \\
  & 6039 & 0.99,-0.02,0.00 & 6918  & 1.00, 0.00 \\ \hline
2 & 132 & 0.02,0.99,-0.02 & 2479 & 0.00,1.00 \\
  & 1326 & 0.00,0.02,0.99 & & \\
  & 4822 & 0.99,-0.02,0.00 & 6216 & 1.00,0.00 \\ \hline
3 & 130 & 0.02,0.99,-0.05 & 2238 & 0.00,1.00 \\
  & 967 & 0.00,0.05,0.99 & & \\
  & 4342 & 0.99,-0.17,0.00 & 5990 & 1.00,0.00 \\ \hline
4 & 119 & 0.02,0.99,-0.09 & 2099 & 0.00,1.00 \\
  & 774 & 0.00,-0.9,0.99 & & \\
  & 4297 & 0.99,-0.02,0.00 & 6394 & 1.00,0.00 \\ \hline
5 & 133 & 0.02,0.99,-0.04 & 2275 & 0.00,1.00 \\
  & 1031 & 0.00,-0.04,0.99 & & \\
  & 5116 & 0.99,-0.02,0.00 & 6800 & 1.00, 0.00 \\ \hline
6 & 128 & 0.02,0.99,-0.07 & 2153 & 0.00,1.00 \\
  & 857 & 0.00,0.07,0.99 & & \\
  & 4823 & 0.99,-0.02,0.00 & 6636 & 1.00, 0.00 \\ \hline
7 & 119 & 0.02,0.99,-0.01 & 2067 & 0.00,1.00 \\
  & 739 & 0.00,0.10,0.99 & & \\
  & 4644 & 0.99,-0.02,0.00 & 6557 & 1.00,0.00 \\ \hline
\end{array}
$$
\end{small}
$$
{\rm Table \ 3}
$$

\end{document}